\documentclass[journal=jpclcd,manuscript=letter]{achemso}

\usepackage{chemformula} 
\usepackage[T1]{fontenc} 
\usepackage{float}
\usepackage{subfig}
\usepackage{pdfpages}

\author{David Ribar}
\affiliation{Computational Chemistry, Lund University, P.O.Box 124, S-221 00 Lund, Sweden}
\author{Clifford E. Woodward}
\affiliation{School of Physical, Environmental and Mathematical Sciences University College, University of New South Wales, ADFA Canberra ACT 2600, Australia}
\author{Jan Forsman}
\email{jan.forsman@compchem.lu.se}
\affiliation{Computational Chemistry, Lund University, P.O.Box 124, S-221 00 Lund, Sweden}

\title{Exceptionally strong double-layer barriers generated by polyampholyte salt}


\begin{document}


\begin{tocentry}
\includegraphics[scale=1]{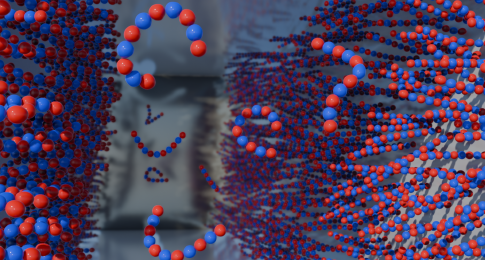}  
\end{tocentry}

\begin{abstract}
  Experiments using the Surface Force Apparatus (SFA) have found anomalously long-ranged interactions between charged
  surfaces in concentrated salt
  solutions. Ion clustering have been suggested as a possible origin of this behaviour.
  In this work, we demonstrate that if such stable clusters indeed form, they are able to induce remarkably
  strong free energy barriers, under conditions where a corresponding solution of simple salt
  provide negligible forces. Our cluster model is based on connected ions producing a polyampholyte salt, containing
  a symmetric mixture of monovalent cationic and anionic polyampholytes.
  Ion distributions and surface interactions are evaluated utilising
  statistical-mechanical ({\em classical}) polymer Density Functional Theory, cDFT. 
  In the Supporting Information, we briefly investigate a range of different polymer architectures (connectivities), but in the main part of the
  work a polyampholyte ion is modelled as a linear chain with alternating charges, in which the
  ends carry an identical charge (hence, a monovalent net charge).
  These salts are able to generate repulsions, between similarly charged surfaces, of a remarkable strength - exceeding
  those from simple salts by orders of magnitude. The underlying mechanism for this is the formation
  of brush-like layers at the surfaces, i.e. the repulsion is strongly related
  to excluded volume effects, in a manner similar to the interaction between surfaces carrying grafted polymers.
  We believe our results are relevant not only to possible mechanisms underlying
  anomalously long-ranged underscreening in concentrated simple salt solutions, but also
  for the potential use of synthesised polyampholyte salt as extremely efficient stabilisers
  of colloidal dispersions. 
  \end{abstract}

\section{}

Using simple salt to regulate the stability of colloidal dispersions that contain charged particles is a
well established process, with a firm theoretical foundation\cite{Derjaguin:41,Verwey:48,Israelachvili:91,Evans:94,Holm:01}.
Mean-field treatments of aqueous systems tends to be accurate at low and intermediate concentrations when the
salts are composed of monovalent species. An important mean-field result is the so-called
Debye screening length, $\lambda_D$, that describes the effective range of electrostatic interactions
in the presence of salt. As the salt concentration increases, {\em ionic screening} leads to a reduced
range, with $\lambda_D \sim 1/c^{1/2}$. However, recent experiments using
the Surface Force Apparatus, SFA, indicate a peculiar non-monotonic behaviour: above
a threshold concentration (usually around 1M), the range of electrostatic forces {\em increase}
upon the addition of salt\cite{Gebbie:13,Gebbie:15,Smith:16,Fung:23}.
This remarkable response has some support from colloidal stability\cite{Yuan:22} and
thin film\cite{Gaddam:19} investigations, but there are
also contradictory experiments\cite{Kumar:22}. Several theoretical efforts
have been made to establish possible underlying molecular
mechanisms \cite{Lee:17,Coupette:18,Rotenberg:18,Kjellander:20,Coles:20,Zeman:20,Cats:21,Hartel:23,Elliot:24,Ribar:24b}
but there is still no broadly accepted view on the matter.

Ion pairing, or ion clusters, have been suggested to play a role for the alleged strong and long-ranged
repulsion at high ionic strengths \cite{Gebbie:13,Ma:15,Hartel:23,Komori:23,Elliot:24,Ribar:24,Ribar:24b}.
In this work, we investigate surface forces in the presence of aqueous solutions that contain
{\em constructed} ``ion clusters'', or at least molecular architectures that can be viewed as simple models
of monovalent ion clusters. Our description is based on an implicit treatment of water, which only enters via its dielectric
constant, $\epsilon_r=78.3$, and ions that are connected to form ``ion clusters'' with a
univalent net charge. This system will be treated using {\em classical} polymer Density Functional Theory, cDFT.

We investigate different polyampholytic architectures, but they are all composed of
connected charged hard spheres with diameter, $d$. The bond between neighbouring charges
in a cluster has a fixed length $b$, and allows for full rotationally flexibility.
The solution
is modelled using an implicit solvent,
and a salt composed of an equimolar mixture of monovalent polyampholyte cations and anions.
The number of charged monomers per chain is denoted by $r$, with each  $r$-mer carrying a net
single elementary charge. In the main article
we will consider a simple linear architecture with alternating charges along the chain.
Thus, the polyampholytic cation will have positive end monomers, and a total of
$(r+1)/2$ positive and $(r-1)/2$ negative charges, with the opposite true for anions. 
Here, we will compare surface interactions in the presence of  a purely polyampholyte salt
with those obtained with a simple salt. 

In the Supporting Information (SI) we explore different
chain architectures and charge distributions. For example, we investigate the 
situation where the charges are collected in blocks 
of monomers of the same valency. Such chains are
likely adopt a folded state, unless there is additional simple salt.
These architectures appear to give rise to rather dramatic effects, but we postpone 
their more detailed study for future work.

Non-bonded monomer-monomer interactions are denoted by  $\phi_{ij}(r)$,  where: 
\begin{equation}
		\beta\phi_{ij}(r) = \left\{
	\begin{array}{ll}
		\infty ; & r \leq d \\
		l_B \frac{Z_i Z_j}{r}; & r > d  \\
	\end{array}
	\right.
	\label{eq:coulomb}
\end{equation}
Here, $\beta = (k_BT)^{-1}$ is the inverse thermal energy, $l_B\approx7.16${\AA} is the Bjerrum length and
$Z_i$ and $Z_j$ denote the valencies of interacting monomers $i$ and $j$ ($|Z|=1$ in all cases).
Monomer densities are denoted $n_+$ (cations) and $n_-$ (anions).

We study the interactions between two flat, hard and infinitely large surfaces, carrying a surface
charge density $\sigma =-1/70 e/${\AA}$^2$, where $e$ is the elementary charge. The surfaces are
immersed in our model mixture and the separation $h$ between the surfaces is varied.
The system is treated using the grand canonical ensemble, i.e. the confined fluid is in chemical equilibrium
with an infinite bulk. The equilibrium  (minimal) free energy is obtained at each 
separation and the net osmotic pressure is evaluated either as a discrete free energy
derivative or from the monomer contact values at the surfaces. All polymer configurations
are accounted for \cite{Woodward:1991}, subject to a Boltzmann weight, under the assumption
that densities only vary in the direction transverse to the surfaces.

For a completely ideal bulk polymer solution, 
in which there are no particle-particle interactions or external fields, save
the bond constraints, the polymer free energy, ${\cal F}_p^{id}$, for an $r-mer$
can be {\em exactly} written as:
\begin{equation}
\beta {\cal F}_p^{(id)}  =  \int N({\bf R}) \left(\ln [N({\bf R})] - 1 \right) d{\bf R} + \beta \int N({\bf R}) V_b({\bf R}) d{\bf R} 
\label{eq:idpol}
\end{equation}
where a polymer configuration is represented by
${\bf R} = ({\bf r}_1,...,{\bf r}_{r})$, and the density distribution 
$N({\bf R})$ is defined such that $N({\bf R})d{\bf R}$ is 
the number of polymer molecules having
configurations between ${\bf R}$ and ${\bf R}+d{\bf R}$. $ V_b({\bf R})$ is the bond potential
between connected monomers. In this work, we will only consider
bonds of fixed length, i.e. $e^{-\beta V_b({\bf R})} \propto \prod \delta(|{\bf r}_{i+1}-{\bf r}_i|-b)$
where $b$ is the bond length, and $\delta (x)$ is the 
Dirac delta function.

For our model of a bulk polyampholyte salt, the Helmholtz free energy is given by:
\begin{equation}
{\cal F}(V,T,N_+,N_-) = \sum_{i=\pm}{\cal F}_i^{(id)} +{\cal F}_{HS}^{(ex)}(n_+,n_-) + {\cal U} 
\label{eq:grandpot}
\end{equation}
We have used the index $i$ to indicate that there are two types of (non-ideal) chains with alternating
charges, one of which (the cation) starts and ends with a positive charge. Anions will have negative ends.
The {\em ideal} contribution to the free energy in the bulk is identical for both species, but in a heterogeneous
environment they will be different. Excluded volume interactions 
are approximated by ${\cal F}_{HS}^{(ex)}$, and is assumed to depend only on total cationic and anionic monomer
density $n_+,n_-$. We have used the  Generalised Flory-Dimer theory to estimate this term \cite{Wichert:96,Forsman:04b}.
All electrostatic interactions are collected in ${\cal U}$, and treated at the mean-field level. 
In our slit environment, we use the grand potential $\Omega$, which can be expressed using the following Legendre transformation
\begin{equation}
  \Omega(V,T,\mu_+,\mu_-;[n_+,n_-,\sigma]) = {\cal F}(V,T;[n_+,n_-,\sigma,\Psi_D]) + A\sum_{i=\pm}\int (V_{ex}(z)-\mu_i)dz
\end{equation}
Here, $z$ is the direction normal to the surfaces, $A$ is the surface area, and $\Psi_D$ is a Donnan potential that
ensures that the system is electroneutral. Polymer configurational entropy, excluded volume and
electrostatic mean-field interactions (including a wall-wall repulsion) are all
contained in ${\cal F}$. The cation/anion chemical potential is denoted by $\mu_+/\mu_-$.
$V_{ex}$ represents the non-electrostatic interaction
with the confining surfaces and, as mentioned, these are purely steric in nature, so
$V_{ex}(z) = 0$ if $d/2<z<(h-d/2)$ and infinite elsewhere.
The grand potential was minimised numerically at each separation $h$, to its equilibrium value, $\Omega_{eq}(h)$, using
Picard iterations. Defining $g_s(h) \equiv \Omega_{eq}/A-p_bh$, where $p_b$ is the bulk pressure, we arrive
at the {\em net} interaction free energy as $\Delta g_s\equiv g_s(h)-g_s(h -> \infty)$.

In order to compare with published SFA data, we will
utilise the Derjaguin Approximation, and transform calculated net free energies per unit area to
the force per radius, $F/R$, appropriate to the crossed cylinder setup in experiments.  In that case,
$F/R = 2\pi \Delta g_s$. The net pressure acting perpendicularly to the flat
surfaces is denoted $p_{net}$, with $p_{net} = -\partial \Delta g_s/\partial h$.

\begin{figure}[h!]
  \centering
  \subfloat[Force per radius, between crossed cylinders.]{
    	\includegraphics[scale=0.36]{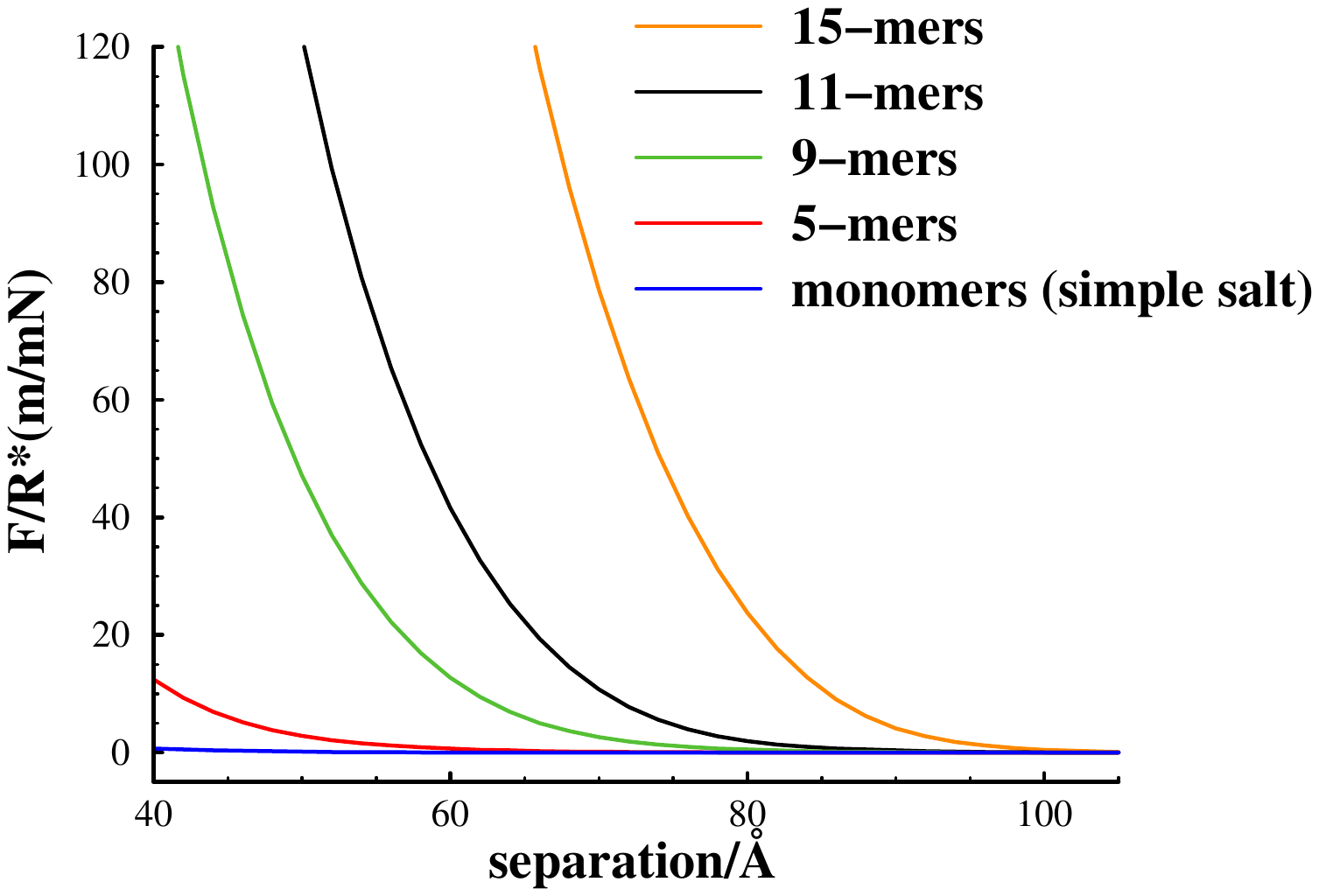}
    }
    \hfill
    \subfloat[The logarithm of the net pressure.]{
	\includegraphics[scale=0.36]{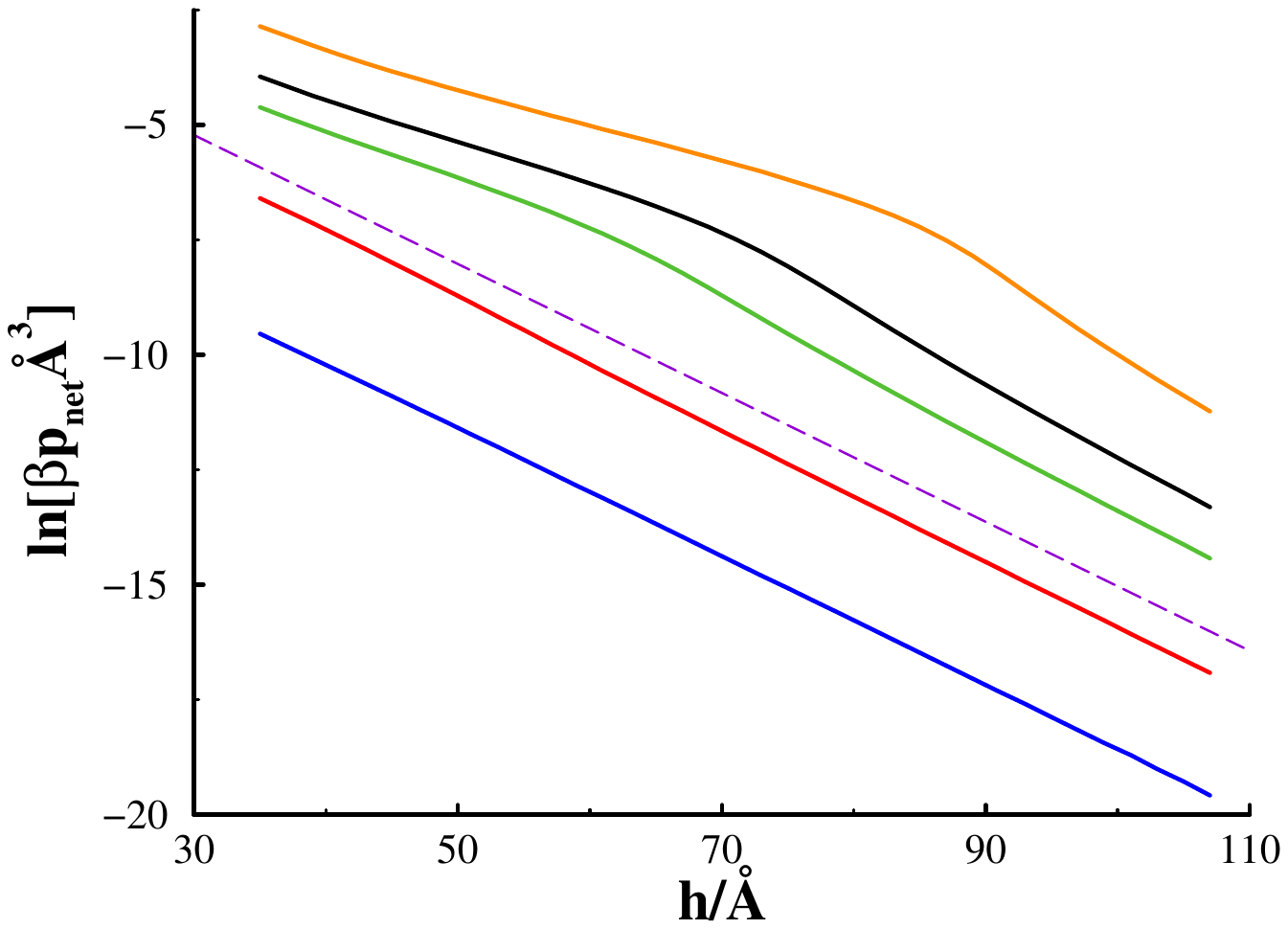}                      
    }
    \caption{Surface interactions at 182mM salt concentration, for simple salts and
      polyampholytes of various lengths. Note that the concentration of charged
      monomers increase with the polymer length. For instance, with 11-mers, the
      bulk concentration of cationic as well as anionic monomers is 2M, but the
      polyampholyte {\em salt} concentration is still 182mM. Legend colours are the same in
      graph (b) as in (a). The thin dashed line in graph (b) shows a line with
      slope $-1/\lambda_D(182mM)$, where $\lambda_D(182mM)$ is the Debye estimate of the screening
      length (about 7.1{\AA}), for a 182mM aqueous solution, with a monovalent salt. The
      placement along the $y$-direction of this dashed line is arbitrarily chosen.}
	\label{fig:polam}
\end{figure}
In Figure \ref{fig:polam}, we compare surface interactions for simple salt, and polyampholyte
salt, at a bulk salt concentration of 182mM. In the latter case, we consider
monodisperse chains of increasing degree of polymerisation, while the chain 
concentration remains fixed at 182mM in the bulk.  Such a scenario  could serve as a
crude approximation to a hypothetical system where ionic clusters begin to form 
in a simple monomeric salt (at 182mM) and any added salt only serves to increase cluster size.
This type of behaviour was actually observed in recent all-atomistic MD simulations
by Komori and Terao\cite{Komori:23}.  They found that increasing
the concentration of monovalent simple salts beyond a critical value
(in that case 1M) gave only an insignificant increase in the free ion concentration.

In Figure \ref{fig:polam}(a), we see a significant increase of the repulsive interaction between
the charged surfaces, compared to monomeric salt, even for rather modest cluster sizes. While the polyampholyte
chains are are assumed to be linear, in the SI we report that quite similar results are
obtain with a branched (star-like) architecture. Despite the growth in 
repulsion with chain length, the long-range decay of the
interaction remains relatively constant. In Figure \ref{fig:polam}(b), we see
that the long-ranged decay is very similar in all cases. For the case of the charged monomers (simple salt), the asymptotic
decay appears to
be exponential, with a decay length that agrees exactly with the Debye length.  This is to be expected as the {\em c}DFT 
uses mean-field electrostatics.  On the other hand, there is a slight but noticeable
decrease in the asymptotic decay length for the polyampholyte systems, with
the decay length decreasing with cluster size.  This is due to the cooperative adsorption 
of chains at the charged surface, which leads to more efficient screening of the surface charge
compared with monomeric salt ions.  
We also notice that at intermediate range polyampholyte-mediated repulsion decays more slowly than with simple salt.
The reason for this will be discussed below.

\begin{figure}[h!]
  \centering
  \subfloat[15-mer monovalent polyampholyte salt.]{
    	\includegraphics[scale=0.36]{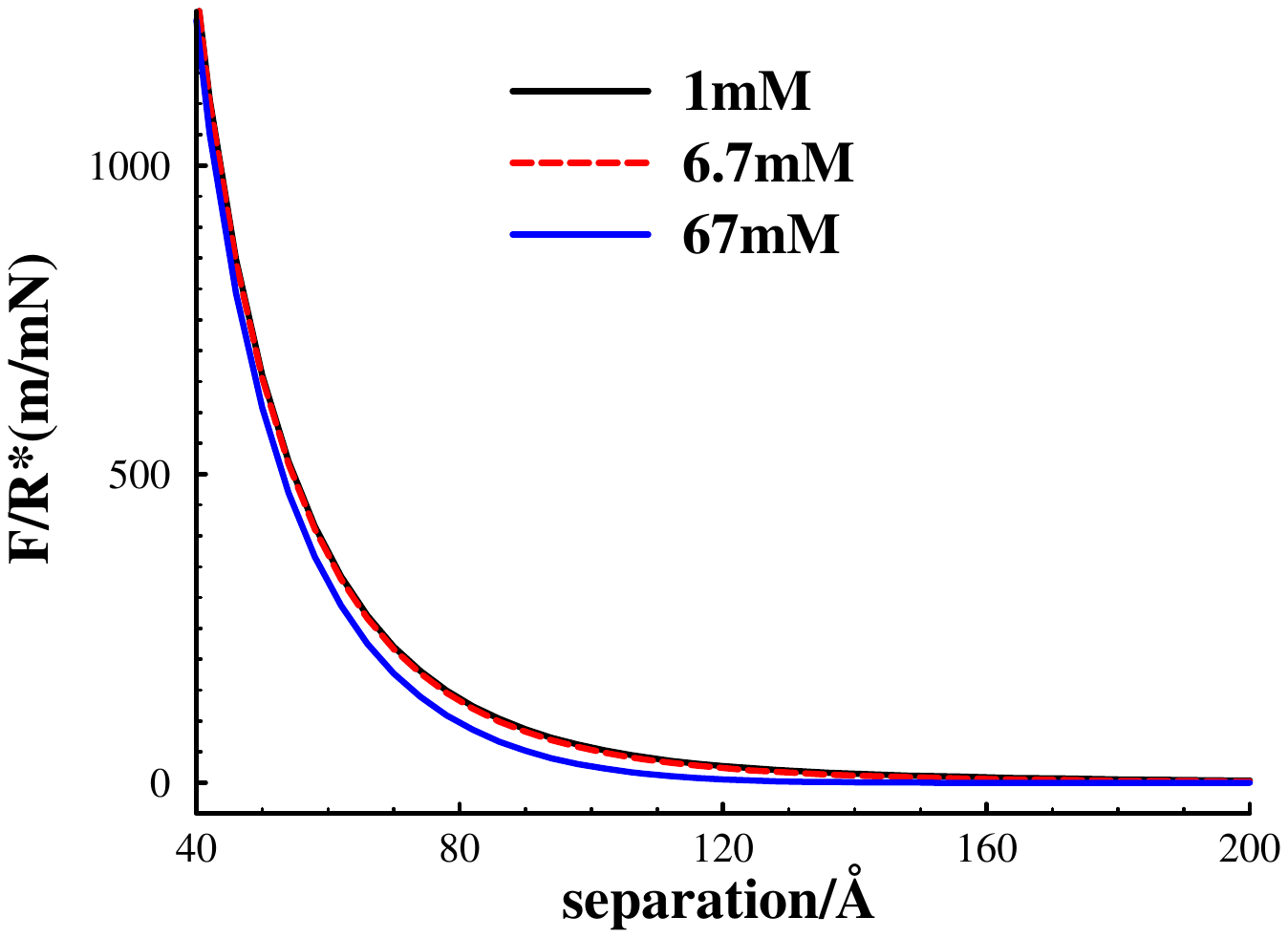}
    }
    \hfill
    \subfloat[Simple 1:1 salt.]{
	\includegraphics[scale=0.36]{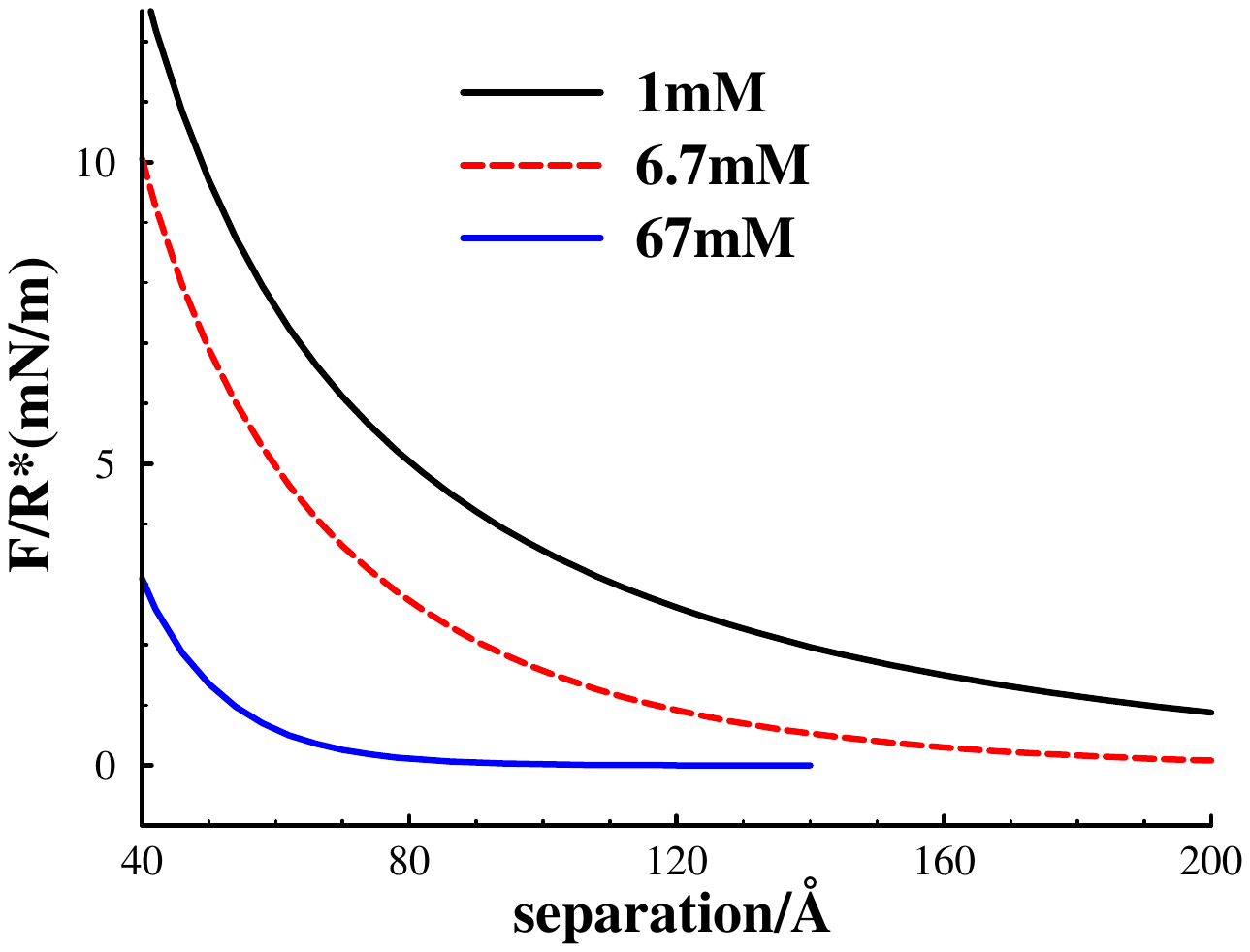}                      
    }
    \caption{Concentration dependence. Interaction free energies
      for polyampholyte (graph(a)) and simple (graph(b)) salts.
      Note the difference in scale between graphs (a) and (b).
    }
	\label{fig:cdep}
\end{figure}
Another interesting property of polyampholyte salt systems is a
remarkable insensitivity to changes of the concentration at
short and intermediate separations. This is shown in graph Figure \ref{fig:cdep} (a), where
we see how a 67-fold increase of the polyampholyte
concentration leaves the surface interactions almost unchanged. 
This is in stark contrast to the response of simple
salt solutions, illustrated in Figure \ref{fig:cdep} (b), where 
the same concentration increase almost eliminates the repulsion.
Note the two order of magnitude difference in scale for the displayed interaction curves.  
\begin{figure}[h!]
  \centering
  \subfloat[15-mer monovalent polyampholyte salt.]{
    	\includegraphics[scale=0.36]{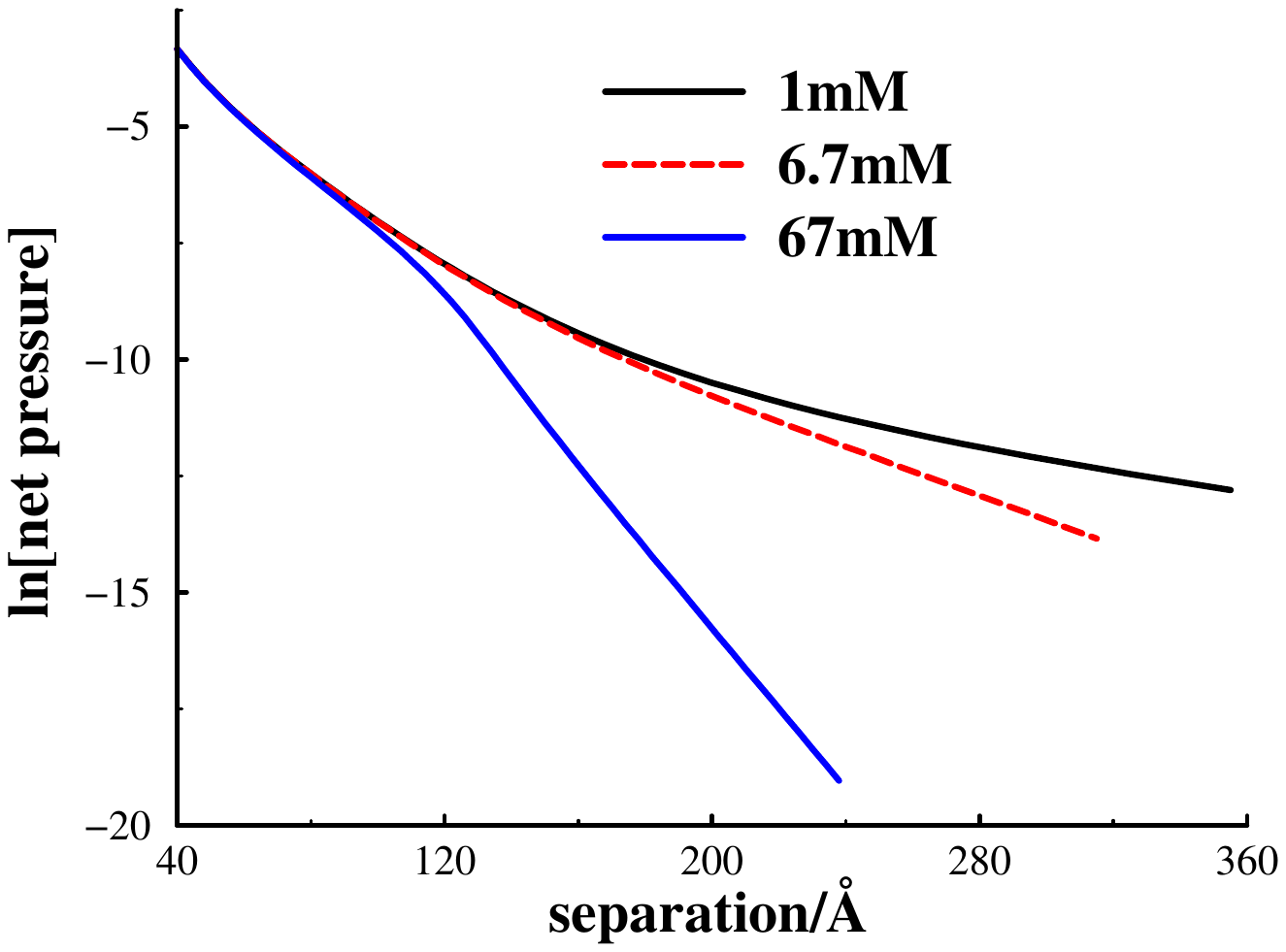}
    }
    \hfill
    \subfloat[Simple 1:1 salt.]{
	\includegraphics[scale=0.36]{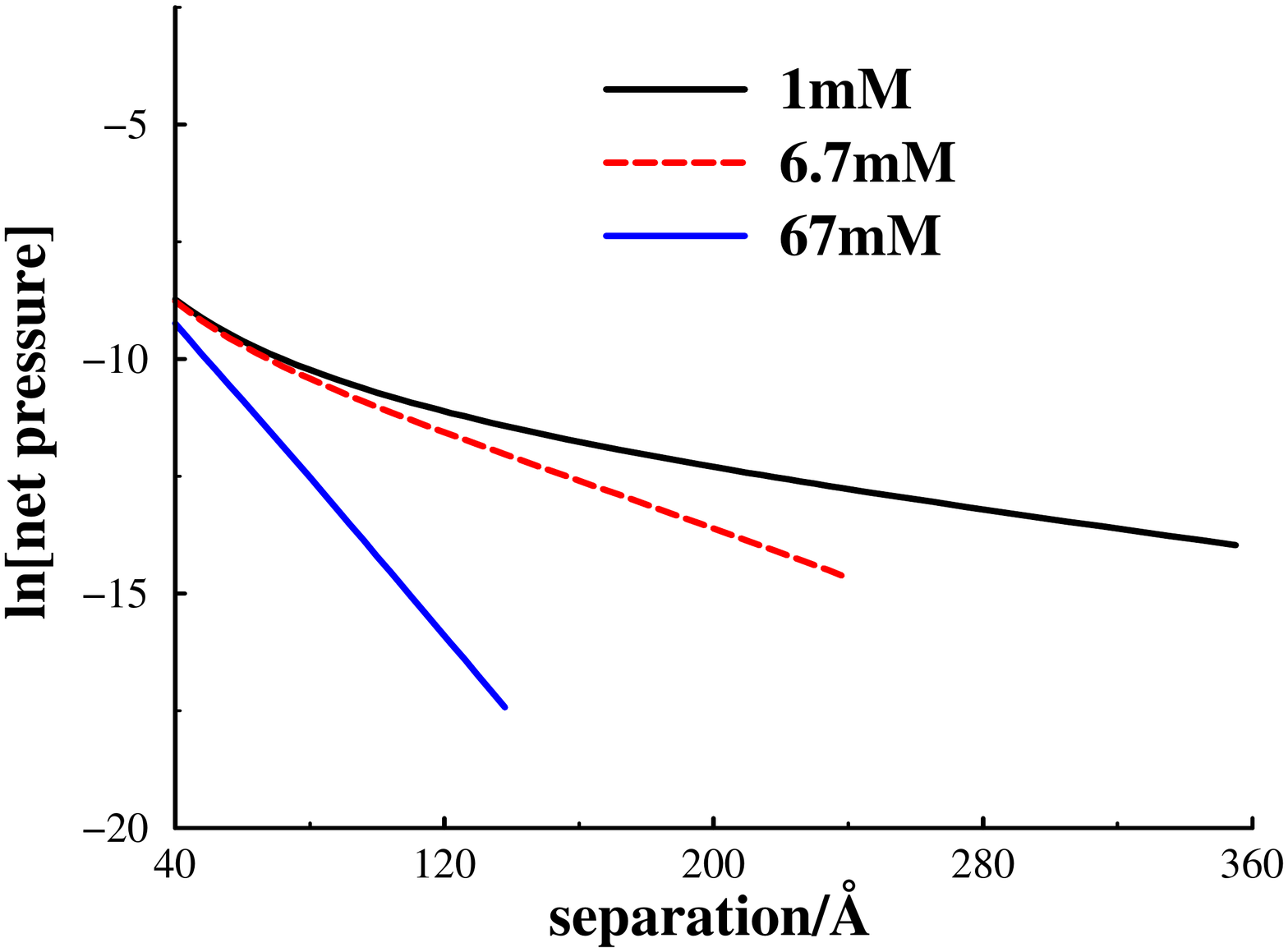}                      
    }
    \caption{Concentration dependence of $\ln{[p_{net}]}$, for polyampholyte (graph(a)) and simple (graph(b)) salts.
    }
	\label{fig:cdep_lnp}
\end{figure}
This is does not mean that the long-range decay length at large separations
is insensitive to the concentration of polyampholyte salt though.
In Figure \ref{fig:cdep_lnp} (a), we see that an increase in the polyampholyte salt
concentration leads to a considerably faster decay at {\em large} separations. As noted above, comparisons
show that the {\em long-ranged} decay length is very similar for polyampholyte and simple salts, Figure \ref{fig:cdep_lnp} (b). 


\begin{figure}[h!]
  \centering
  \subfloat[Interaction free energies, between charged surfaces immersed in a 1mM 15-mer (monovalent) polyampholyte salt.
  Results are shown for polymers with monomers of sizes $d=0->4${\AA}.]{
    	\includegraphics[scale=0.36]{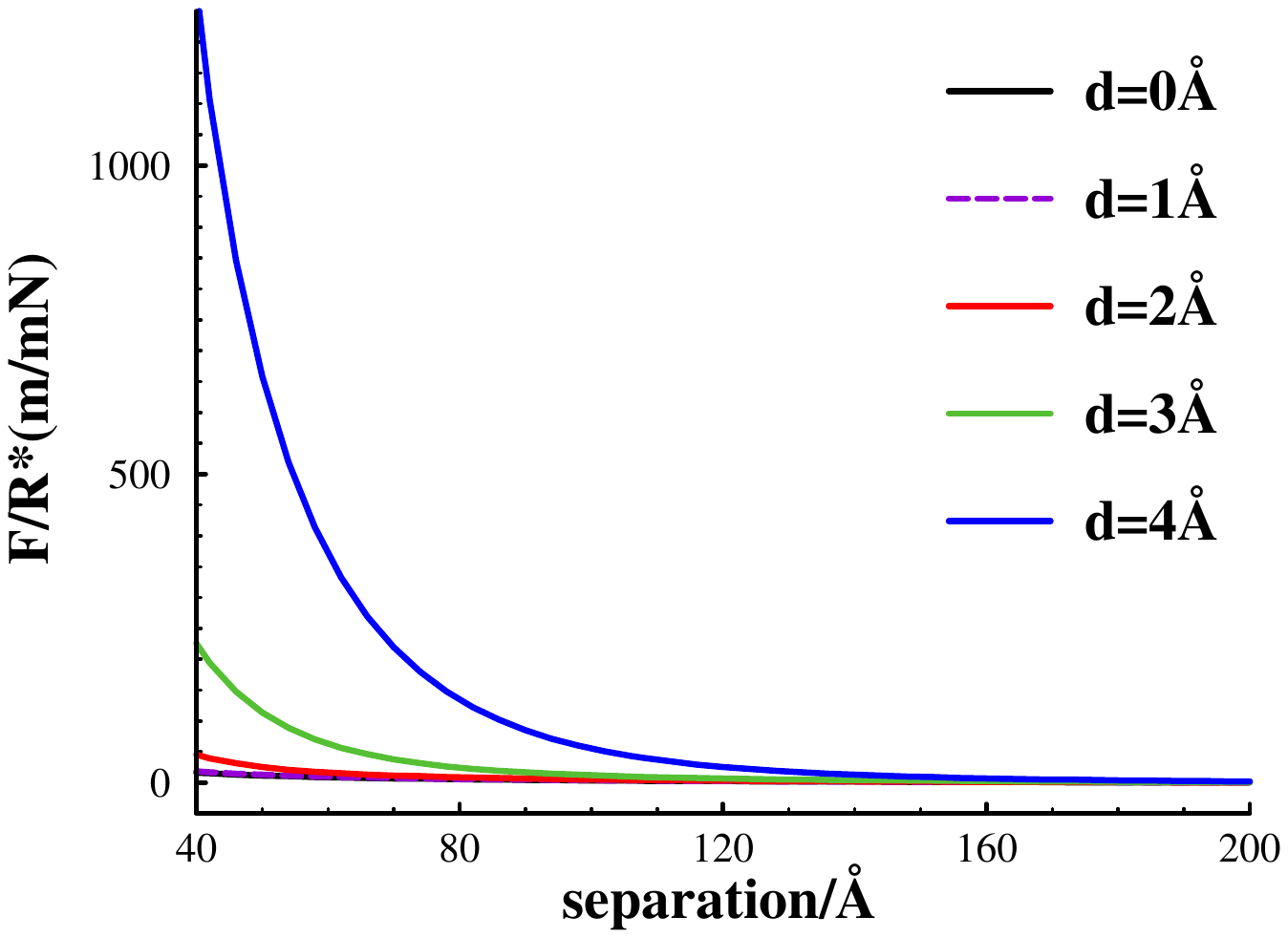}
    }
    \hfill
    \subfloat[Cationic monomer density profiles at one of the charged surfaces, for the same systems as in graph (a).
    The surface separation is 108\AA in all cases.]{
	\includegraphics[scale=0.36]{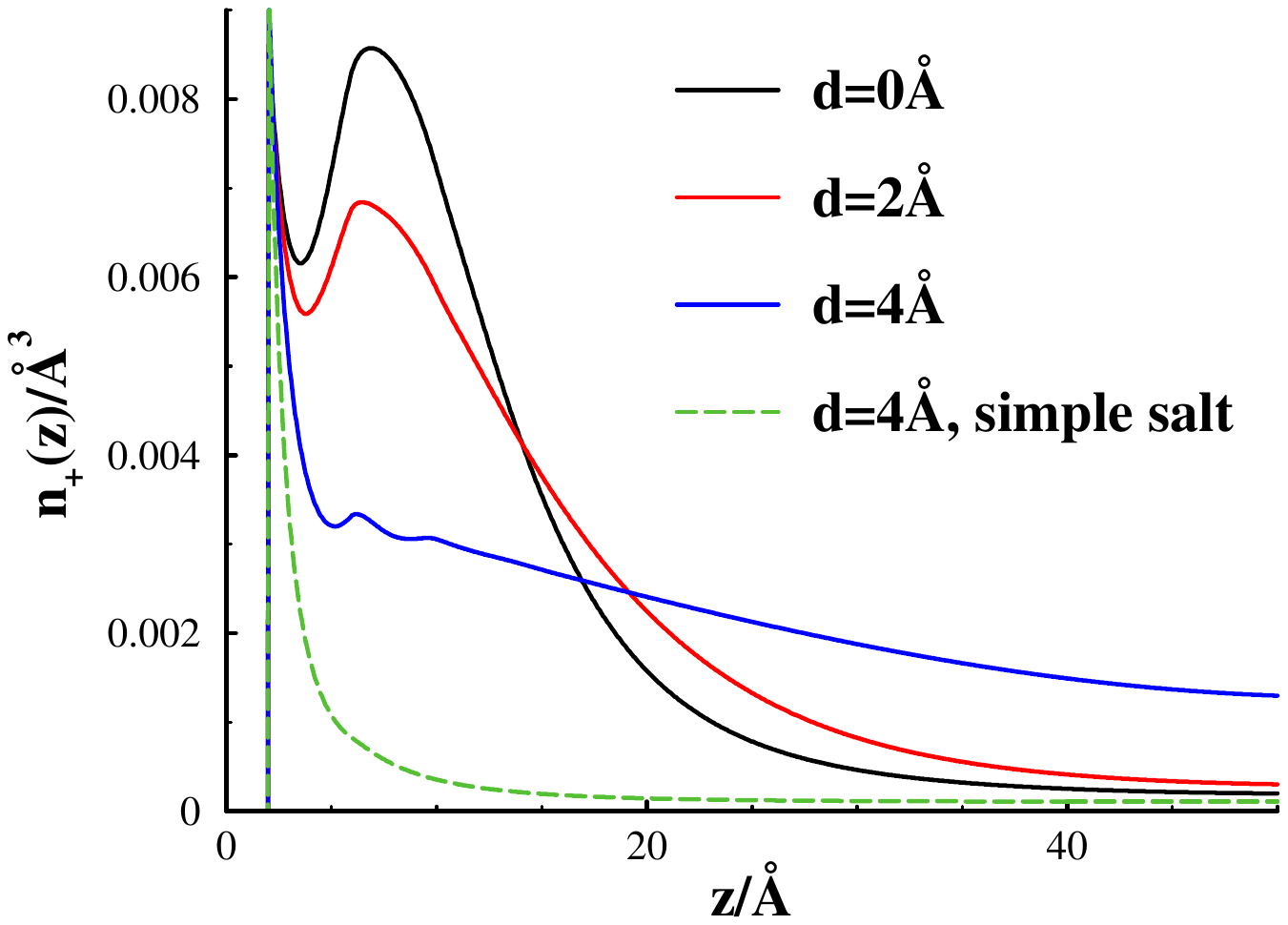}                      
    }
    \caption{Excluded volume effects, evaluated by changes to the monomer hard-sphere diameter.
    }
	\label{fig:exclvol}
\end{figure}
So, where does the very strong repulsion at intermediate separations
seen in the polyampholyte salts originate from? 
It turns out that steric interactions play a crucial role. 
This is exemplified in Figure \ref{fig:exclvol} (a), where we note
that a reduction of the monomer hard-sphere diameter leads to a dramatic drop of
the surface force. This is in turn related to the monomer density distribution at the surfaces, as shown
in Figure \ref{fig:exclvol} (b). Only cationic monomer densities are displayed, but the overall
trend is the same for anions. As the monomer radii are decreased, the chains adopt a much 
more compact density profile.  This suggests weaker chain-chain overlap  as the surfaces are 
brought closer together.  Thus, we conclude that the strong intermediate repulsions 
are due to chain-chain overlaps dominated by steric interactions.  
In this context we note that the concentration of monomeric (simple) salt at the surfaces
is generally much smaller than that of the polyampholyte monomers, leading to 
very weak repulsive interactions, by comparison.  The
apparent insensitivity  of this intermediate repulsion on the concentration
of the polyampholyte chains means that the surface adsorption of the 
chains reaches saturation at fairly low concentrations.

The mechanism for this intermediate separation regime is similar to that between surfaces with adsorbed polymer layers.
In that case we know that the range of the surface interactions is determined by the chain length.   So why is this
interesting?  In Figure \ref{fig:polam}(b), we saw that chain-chain interactions
gave rise to an intermediate decay length which was longer than the Debye length.
Though our results show that the Debye length  (albeit slightly modified) was relevant at very large surface separations, 
our results may still have relevance to the anomalous interactions found in SFA measurements in simple
concentrated salts.  In this work, we have limited ourselves to  chains of finite length.
Thus at very large surface separations, the overlaps between finite chains are 
less important compared to electrostatic effects, and this is why the usual Debye-like 
decay of the interactions eventually prevail asymptotically. For aggregating clusters, the
chain size distribution will likely have an exponential form and in this case,
chain overlaps remain significant at much larger surface separations, possibly
dominating the electrostatic mechanism asymptotically and giving rise to apparently anomalous
long-range decays.

We envisage that our theoretical predictions are readily verifiable by surface force measurements, for instance
using polypeptides that are composed of alternating amino acids of opposite charge (at some suitable pH).
Upon verification, such polyampholyte salt systems could be further explored in the context of
colloidal stability.

\begin{acknowledgement}
  J.F. acknowledges financial support by the Swedish Research Council, and computational resources by the
  Lund University computer cluster organisation, LUNARC. We also thank prof. Sture Nordholm for
  fruitful discussions. 

\end{acknowledgement}

\begin{suppinfo}

The following files are available free of charge.
\begin{itemize}
  \item Supporting information: Detailed simulation methods, and further analyses.
  \item Github repository: all codes used for simulations, along with the data generated, is freely available.
\end{itemize}

\end{suppinfo}

\bibliography{poly.bib}



\section*{Supplementary material}

# Supporting information for:

# Exceptionally strong double-layer barriers generated by polyampholyte salt

David Ribar[a], Clifford E. Woodward[b], Jan Forsman[a,*]

[a]*Computational Chemistry, Lund University, P.O.Box 124, S-221 00 Lund, Sweden*

[b]*School of Physical, Environmental and Mathematical Sciences University College, University of New South Wales, ADFA Canberra ACT 2600, Australia*

*Corresponding author: jan.forsman@compchem.lu.se

(Dated: 2024-12-04)

# S1 Model illustration

We have constructed a cartoon illustration of our polyampholyte salt model. It is shown, for 5-mer

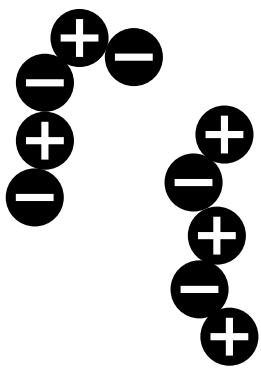

Figure 1: A graphical illustration of a cation and an anion, in a 5-mer polyampholyte salt.

polyampholytes, in Figure 1. This is for a case where the hard-sphere diameter $d$ equals the bond length $b$, i.e. $d = b = 4$Å. In the main paper, we also considered cases with smaller monomers.

# S2 Other polymer architectures

Here we will briefly consider other modelling options for the polyampholytes, which we recall also might serve as a crude description of ion clusters. One might anticipate a rather compact arrangement of an ion cluster, and a step in that direction would be to allow the polymers to become star-like, with several branches connecting to a common central monomer. However, we see in Figure 2 that the resulting interaction free energies are quite similar for branched and linear polymer architectures. Moreover, distributing the single net charge per chain uniformly among the monomers leads to quite similar results as for monovalent polympholytes in which the monomers have alternating charge. On the other hand, if the charges are collected into two separate blocks with opposing valency (one block containing one more monomer than the other), then the repulsive force become much stronger. One should then keep in mind that the construction of such polymers in a "real" solution might lead to folded structures, at least for

1

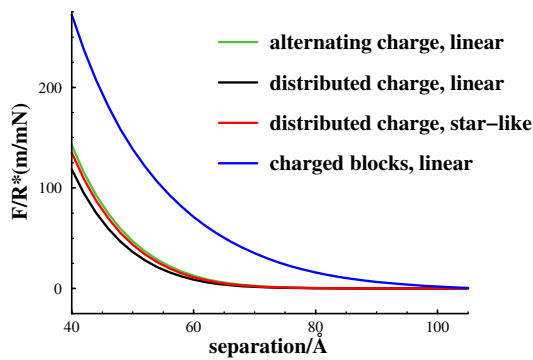

Figure 2: Interactions between charged surfaces, immersed in a solution containing 182mM 9-mer polyampholyte salts. The green curve are results using our reference model (main paper), i.e. composed of monomers with alternating charge. We also display (black) results using linearly connected chains with a single charge that is equally distributed between the monomers ($\pm e/9$ charge per monomer, resulting in monovalent chains). In the latter case, we furthermore show (red) corresponding interactions when the polymers have a 4-armed star structure, i.e. 4 branches connecting to a common central monomer. Finally, the blue curve shows interactions in the presence of polymers with a linear block charge structure ($+ + + + + - - - -$ and $+ + + + - - - - -$).

large degrees of polymerisation. The tendency to fold can be counteracted by the addition of simple salt, and interactions in the presence of such salt mixtures will be the focus of a future work.

Let us finally illustrate the relevance of excluded volume interactions, also when the monovalent charge is evenly distributed among the monomers, and when the polymers have a star-like structure. This is

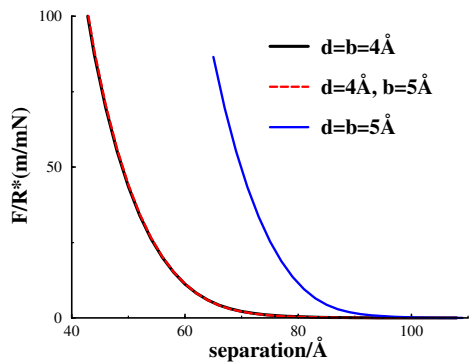

Figure 3: Interactions between charged surfaces, immersed in a solution containing 182mM 9-mer polyampholyte salts. Here, we consider chains with a 4-armed star structure, and various choices for the bond length ($b$) and monomer hard-sphere diameter ($d$).

illustrated in Figure 3, under conditions similar to those in Figure 2, i.e. 182mM monovalent 9-mer salt. We note that for a given monomer diameter, the bond length has a weak influence. On the other hand, increasing the hard-sphere monomer diameter generates a dramatically stronger repulsion.

2



\end{document}